\documentclass[final,3p]{elsarticle}

\usepackage{graphicx}

\usepackage{amssymb}
\usepackage{lineno}
\usepackage{lscape}
\usepackage{tabularx}
\usepackage[gen]{eurosym}
\usepackage{multirow}
\usepackage{hhline}
\usepackage{booktabs}
\usepackage{longtable}
\usepackage{siunitx}
\usepackage{framed}
\usepackage{todonotes}
\usepackage{wasysym}

\usepackage{pbox}
\usepackage{float}
\usepackage{framed}

\journal{The Journal of Software: Evolution and Process (Accepted Nov 16, 2019)}

\begin{document}

\begin{frontmatter}

\title{Agile Ways of Working: A Team Maturity Perspective}

\author[chalmersgu]{Lucas Gren\corref{cor1}}
\ead{lucas.gren@cse.gu.se}
\cortext[cor1]{Corresponding author. Tel.: +46 739 882 010}

\author[usp]{Alfredo Goldman}
\ead{gold@ime.usp.br}

\author[psy]{Christian Jacobsson}
\ead{christian.jacobsson@gu.se}

\address[chalmersgu]{Chalmers University of Technology and the University of Gothenburg, Gothenburg, Sweden}

\address[usp]{The Department of Mathematics and Statistics, The University of S\~ao Paulo, S\~ao Paulo, Brazil}

\address[psy]{The Department of Psychology, The University of Gothenburg, Gothenburg, Sweden}

\begin{abstract}
With the agile approach to managing software development projects comes an increased dependability on well functioning teams, since many of the practices are built on teamwork. The objective of this study was to investigate if, and how, team development from a group psychological perspective is related to some work practices of agile teams. Data were collected from 34 agile teams (200 individuals) from six software development organizations and one university in both Brazil and Sweden using the Group Development Questionnaire (Scale IV) and the Perceptive Agile Measurement (PAM). The result indicates a strong correlation between levels of group maturity and the two agile practices \emph{iterative development} and \emph{retrospectives}. We, therefore, conclude that agile teams at different group development stages adopt parts of team agility differently, thus confirming previous studies but with more data and by investigating concrete and applied agile practices. We thereby add evidence to the hypothesis that an agile implementation and management of agile projects need to be adapted to the group maturity levels of the agile teams.
\end{abstract}


\end{frontmatter}

\section{Introduction}
In the context of software engineering, a set of practices is applied to manage complex and rapidly changing projects in relation to their requirements \cite{dingsoyr20121213}. In order to succeed and be responsive to such change, research and practice advocate an increased focus on collaboration \cite{hoda2017systematic}.

The basic idea of agile software development is that complex projects need to combine managing projects with the need to be able to respond to change. A core practice of any agile method is therefore to develop the product iteratively, meaning that the projects are divided into shorter iterations so that the requirements can be re-prioritized continuously. Compared to other management approaches, this is the key difference in agile software development, since it is the only project management method that assumes the end-goal is unknown in detail throughout the project life-cycle \cite{beck1999embracing}. Another key idea about how such responsiveness to change should be possible, is to focus more on social-psychological aspects both within the teams and with customers. Cohen et al. \cite{cohen} even go as far as to say that ``/.../ what is new about Agile Methods is not the practices they use, but their recognition of people as the primary drivers of project success, coupled with an intense focus on effectiveness and maneuverability'' \cite{cohen}. In a secondary study on behavioral software engineering (i.e.\ human and social aspects of software engineering), Lenberg et al.\ \cite{lenberg2015} concluded that studies have mostly focused on a few concepts, which have been applied to a limited number of software engineering areas. Wufka and Ralph \cite{wufka2015explaining} also concluded that agility emerges from teams' reactions to needs for change and that agility needs to be understood from a process (i.e.\ temporal) perspective of teams. 

Therefore, the contribution of this study is an investigation of a subset of the agile practices' correlation to team maturity. This then adds the team maturity concept, from a collaborative perspective, to the understanding of the use of these agile practices in industry and how to improve them. The remaining part of the paper is structured as follows; in the next section (Section~\ref{ork}) we present a background both in connection to research on agile teams in software engineering, but also some studies on teams in the psychology field. We motivate the study and suggest hypotheses in Section~\ref{rg}. Section~\ref{method} presents the method of the study. Section~\ref{sec:res} shows results, and in Section~\ref{disc} we discuss the results, in Section~\ref{vts} threats to validity, and in Section~\ref{conc}, we conclude the paper and suggest future work.

\section{Background}\label{ork}
\subsection{Research on Agile Teams in Software Engineering}
In recent years, the empirical studies on agile software development have increased and focus on different aspects of agile software development and management \cite{hoda2017systematic}. The areas that have been covered by secondary studies are: adoption, methods, practices, human and social aspects, CMMI (an organizational maturity model), usability, global software engineering (GSE), organizational agility, embedded systems, and software product line engineering \cite{hoda2017systematic}. The three systematic literature reviews covering human and social aspects and organizational agility were about dimensions of organizational agility (organizational structures, workforce, development process, management and leadership, and infrastructure) \cite{tapanainen2008towards}, the role of communication \cite{hummel2013role}, and developers' motivation in agile projects \cite{melo2012developers}. From both a research and industry perspective, measuring \emph{agility} is of course of very high value. However, the vagueness and breadth of the agile descriptions make a definition of one single construct of \emph{agility} in its broader sense very difficult to achieve. The research on agile software development has, therefore, naturally become split into different subcategories, like what is meant by \emph{agile requirements engineering} \cite{inayat2015systematic}, \emph{agile contracting} (see e.g.\ Hoda et al.\ \cite{hoda2009negotiating}), and so on and so forth.

In a study by McDonald and Edwards \cite{mcdonald} in the software engineering domain, the authors conclude that as much is derived from the work-group's context, as of the people in it. Still, most studies still focus on individual psychology, such as Seger et al.'s \cite{seger} study on psychological needs connected to software development teams. As a side-note, the term \emph{team} is almost used exclusively for workgroups in software engineering research and practice and denotes a small workgroup at an organization with people that collaborate with each other. As previously mentioned, human factors have gotten more attention in software engineering \cite{bali}. Lenberg et al.\ \cite{lenberg2015} also believe more focus on these factors in research is needed. Feldt et al.\ \cite{feldt2010} also argue for the usage of personality tests to put together teams, even though they state that personality cannot be considered in isolation. An indication of this, is a more recent study by Cruz et al.\ \cite{forty} showing that 40 years of using personality tests in software engineering does not give any coherent results. Furthermore, studies by Hannay et al.\ \cite{hannay2010effects} and Licorish and MacDonell \cite{licorish2015communication} have shown that personality tests have little predictive value and looking at behavior in context could be a better approach if prediction is the goal. We therefore want to use the agile team as the level of analysis in this current study.

Another study was conducted by Moe et al.\ \cite{moe2010}, where they studied a Scrum project that was implemented in an organization. They concluded that transition to self-organizing teams needs buy-in from both developers and managers. Team orientation, team leadership and coordination, as well as division of work, were factors to consider when moving towards becoming an agile team. The same authors also concluded that challenges when implementing shared decision-making were: alignment of strategic product plans with iteration plans, allocation of development resources, and performing development and maintenance tasks in teams \cite{moe2012}. Hoda et al.\ \cite{hoda2013self} also showed the importance of having defined roles in agile teams, which is something stressed as very important in group development research \cite{wheelandev}. The aspects of all these three studies are parts of group development and therefore motivate studying the relationship between the agile ways of working and group, or team, maturity. 

Relating agile software development to psychological theories, also advocated by Lenberg et al.\ \cite{lenberg2015}, could provide a deeper understanding of the psychological processes in the agile workplace. Such understanding would then hopefully lead to better predictability and intervention in relation to human factors in agile projects, but does also increase our understanding of the high capacity of mature groups in relation to dealing with volatile demand. 

\subsection{Agile work practices}\label{agileprac}
There are many agile approaches to developing software, e.g.\ eXtreme Programming (or XP) \cite{beck1999embracing}), Kanban \cite{ahmad2013kanban} and Scrum \cite{rising2000scrum}, which try implement responsiveness to changing requirement in form of concrete practices. These practices are widely used in industry \cite{7890614}, but it is difficult to find scientific studies on their usefulness in relation to concrete benefits for the developed software other than that customers are overall more satisfied with an agile approach \cite{serrador2015does}. Most descriptions are from popular books with no or very little scientific underpinnings (see e.g.\ \cite{sean,cooke,schwaber,weinberg,kim2014phoenix,agilecobb,chin,cockburn,vanOosterhout2010business}), and the scientific studies on scaled agile is still scarce with only five studies in 2018 \cite{paa}. Some of the practices stem from different fields, like Kanban that comes from Lean Production \cite{poth2019lean}. We have only found one study \cite{so2010making} that has explicit connections between some very common agile practices \cite{7890614} and direct team benefits. These practices mostly stem from Scrum and we will not investigate Lean in particular in this paper. The author presents a quite old (2010) but thorough quantitative study on agile practices and goal commitment, social support, adaptation, and knowledge growth \cite{so2010making}. 

We use the eight agile practices suggested by So \cite{so2010making} in this current study. However, we do not consider these the only agile practices that are important for agile teams, we used these eight in this study because we wanted as validated surveys as possible. These practices, or similar variants, are also presented as some of the most common ones in Licorish et al. \cite{7890614}. The practices are: \emph{Iterative development}, \emph{Continuous integration and testing}, \emph{Stand-up meetings}, \emph{Customer acceptance tests}, \emph{Retrospectives}, \emph{Iteration planning}, and \emph{Customer access}.

\paragraph{Iterative development}
The idea of delivering software iteratively is to deliver working software comprising some initial functionality that gives value to the customer at a very early point in time in the development life-cycle. No big planning is therefore conducted up-front, but the next increment is decided with the customer continuously \cite{Jalote200467}. Two example items from the measurement of this practice by So \cite{so2010making} are: 1) When the scope could not be implemented due to constraints, the team held active discussions on re-prioritization with the customer on what to finish within the iteration. 2) Working software was the primary measure for project progress.

\paragraph{Continuous integration and testing}
Continuous integration and testing prescribe a continuous integration of source code into the software covered by different tests \cite{so}. Two example items from the measurement of this practice by So \cite{so2010making} are: 1) New code was written with unit tests covering its main functionality. 2) For detecting bugs, test reports from automated unit tests were systematically used to capture the bugs.

\paragraph{Stand-up meetings}
A common practice to give teams the possibility to coordinate, is the daily stand-up. This meeting is time-boxed (a rule of thumb is around 15 minutes which is the amount of time people can `stand up'), and the three standard questions to be answered by every team members are: (1) What have you done since we last met? (2) What are you planning to do until we meet again? (3) What, if any, impediments are you encountering that are preventing you from making progress? \cite{stray2016daily}. Two example items from the measurement of this practice by So \cite{so2010making} are: 1) Stand up meetings were to the point, focusing only on what had been done and needed to be done on that day. 2) When people reported problems in the stand up meetings, team members offered to help instantly.

\paragraph{Customer acceptance tests}
In a customer acceptance test, the customer gets to see and test the latest working version of the product with some functionality of high priority to provide continuous feedback, but also to be given the possibility to change the requirements of the product \cite{beck1999embracing}. This is done in order to incorporate responsiveness to change. Two example items from the measurement of this practice by So \cite{so2010making} are: 1) A requirement was not regarded as finished until its acceptance tests (with the customer) had passed. 2) The customer provided a comprehensive set of test criteria for customer acceptance.

\paragraph{Retrospectives}
In a retrospective meeting, the team members are to reflect on how their collaboration and work is going and find improvements for the next iteration \cite{derby2006arm}. Two example items from the measurement of this practice by So \cite{so2010making} are: 1) The retrospectives helped us become aware of what we should improve in the upcoming iteration\slash s. 2) Our team followed up intensively on the progress of each improvement point elaborated in a retrospective.

\paragraph{Iteration planning}
The process of planning an iteration is often called the \emph{Planning Game} in the agile context \cite{karlsson2007pair}. When planning an upcoming iteration,  the team (including the customer or a customer representative) prioritizes and conducts effort estimation on user stories \cite{mahnivc2012using}. Two example items from the measurement of this practice by So \cite{so2010making} are: 1) All members of the technical team actively participated during iteration planning meetings. 2) All concerns from team members about reaching the iteration goals were considered.

\paragraph{Customer access}
Customer access is very more important for agile teams since they are to be responsive to changes in the prioritization of what they are to deliver next \cite{krebs2002turning}. Two example items from the measurement of this practice by So \cite{so2010making} are: 1) The developers could contact the customer directly or through a customer contact person without any bureaucratic hurdles. 2) The feedback from the customer was clear and clarified his requirements or open issues to the developers.

\subsection{Research on Teams in Psychology}
Keyton \cite{grupp} defines a group as three or more members that interact with each other to perform a number of tasks and achieve a set of common goals. This means that many large groups are in fact a set of smaller subgroups and should be handled as separate groups. If the group consists of more than eight individuals, they are less productive than smaller groups, and research findings suggest that small work groups of 3 to 6 members have a much better chance of reaching the higher stages of group development than larger groups \cite{wheelan2009}. A \emph{work group} is a composition of members that is striving to create a shared view of goals and trying to develop a structure to achieve these goals.

The study of the behavior of small groups was launched with the establishment of a research center for group dynamics in 1946. Several research groups proposed different ways of analyzing the behavior of groups \cite{wheelandev}. Some studies propose group development can be described as states or levels of activity (cyclic models, see e.g.\ Bion \cite{bion}), but an integrative theory of linear and cyclic models was first introduced in 1964 according to Wheelan \cite{wheelandev}. A large integrative analysis of various group development models was conducted by Tuckman \cite{tuckman1965developmental}. The result of his analysis was a conceptual model including four stages of group development, namely, Forming, Storming, Norming, and Performing. The model suggested by Wheelan \cite{wheelandev} largely overlaps these stages and is presented next.

\paragraph{The Integrated Model of Group Development}\label{sub:integratedgroup}
Wheelan \cite{wheelandev} created an integrated model of group development with four different development stages. 

Stage 1: ``Dependency and Inclusion''. The first stage is categorized by three main areas; concerns about safety and inclusion, member dependency on the designated leader, and a wish for order and structure. The group is supposed to become organized, capable of efficient work, and achieve goals, so the first state must have a purpose in getting there \cite{wheelan}. 

Stage 2: ``Counter-Dependency and Fight''. The second stage of a group's development is a conflict phase where fight is a must in order to create clear roles to be able to work together in a constructive way. The members have to go through this in order to be able to trust each other and the leader. When the group safely navigated through the first stage they have gained a sense of loyalty. As people feel more safety they will dare to speak up and express opinions that might not be shared by all members \cite{wheelan}.

Stage 3: ``Trust and Structure''. The third stage is a structuring phase where the roles are based on competence instead of striving for power or safety. Communication will be more open and task-oriented. The third stage of group development is characterized by more mature negotiations about roles, organization, and processes \cite{wheelan}. 

Stage 4: ``Work and Productivity''. The fourth and final stage (excluding the termination phase) is when the group wants to get the task done well at the same time as the group cohesion is maintained over a long period of time. The group also focuses on decision-making and encourages task-related conflicts. This is a time of intense productivity and effectiveness and it is at this stage the group becomes a team \cite{wheelan}.

The largest contribution of Wheelan and Hochberger \cite{wheelandev} is probably to connect a questionnaire to the suggested model of group development (the Group Development Questionnaire \cite{wheelan}). In doing so, it has become possible to diagnose and pinpoint in what group stage the group focuses its energy, and therefore provides a means to move forward in its development. 

\section{Research gap and hypotheses}\label{rg}
Gren et al.\ \cite{grenjss2} analysed group development aspects qualitatively and concluded that the definitions of agile teams and mature teams overlap. They also used a version of the agile maturity model suggested by Sidky et al.\ \cite{sidky} and showed correlations between Wheelan's \cite{wheelandev} Group Development Questionnaire Scale IV (Work and productivity), and agile organizational behavior. A problem with that study is that they used a modified version of Sidky et al.'s \cite{sidky} survey, which is quite high level with regards to agility and do not accommodate concrete agile practices used. Their study was also conducted on a smaller sample ($N=66$), which warrants a replication with validated scales and larger data sets.

No studies have been found that specifically connect the temporal perspective of group dynamics (i.e.\ group development) and concrete agile software development team practices. The Group Development Questionnaire Scale IV has been shown to be correlated to a set of performance measurements in many different fields, e.g.\ an ability to finish projects faster \cite{wheelan1998}, better student performance on standardized test (SAT scores) if faculty team is mature \cite{wheelan1999,wheelan2005}, and lower mortality rates in intensive care units performing surgery \cite{wheelan20032}.

The agile practices described above are explained in a way that they seem to be tailored for quite mature work-groups from a group developmental perspective \cite{tuckman,wheelan2003}, i.e.\ it is hard to see how they would be implemented in very immature groups, for example. In order to fully understand the social-psychological components of the team-based workplace in general, and the agile context in particular, we also need to investigate the temporal perspective of the interplay between group development and the agile approach to projects. In addition, software engineering provides a context where projects are highly complex and rapidly changing, which is somewhat a new context of research for more general small group research. The existing research on group effectiveness has also been criticized for having been built on a relatively small number of groups from specific contexts \cite{kozlowski2006enhancing}. It is therefore important to apply a temporal perspective to the agile team dynamics, since team agility might be dependent on the maturity of work groups and therefore need different and contextualized implementations. In this present study we, therefore, selected to focus on this perspective and opted to use the group development questionnaire (GDQ), since it is built on the empirically tested idea that groups develop across time \cite{wheelan2003}. It is also important to note that a team that has been stable and met over a longer period of time is more likely to have matured \cite{wheelan2003}, although group can get stuck in the earlier stages \cite{wheelandev}. Even though we used a cross-sectional design, the result of the GDQ has a built-in temporal perspective and is a measurement of where a small group is in its development.

We have only found one study that has summarized a set of agile practices that the authors argue are common to most agile methods, created a scale for each practice, and validated these factors by using psychometrics, which is the study by So and Scholl \cite{so} presented above. In this present study, we therefore investigate the connections between these two measurements. Since we, at this point, do not know the relationship between dependent and independent variables, we chose to conduct a bivariate correlation analysis. We do not imply that these seven agile practices cover the concept of agility by any means, this study is just a first step toward understanding these connections and we only argue that the practices suggested by So and Scholl \cite{so} at least cover some aspect of agile ways of working.

Hence, we set up the research hypothesis that any of the seven agile practices as defined and operationalized by So and Scholl \cite{so} (\emph{iterative development, continuous integration and testing, stand-up meetings, customer acceptance tests, retrospectives, iteration planning, and customer access}) are correlated to \emph{group maturity} as defined and operationalized by Wheelan \cite{wheelan}. The null and the seven alternative hypotheses for each agile practice are:

\begin{itemize}
\item The null hypothesis (H$_0$) in all cases is that the correlation coefficient is zero, i.e.\ that there is no significant correlation between Team Maturity (tm) and the other variable. 

$H_{0}: \rho_{tm,id} = \rho_{tm,cit} = \rho_{tm,sum} = \rho_{tm,cat} = \rho_{tm,r} = \rho_{tm,ip} = \rho_{tm,ca} = 0$
\item H$_1$) There is a significant correlation between Team Maturity and Iterative Development (id).

  $H_{1}:\rho_{tm,id} \neq 0$ 
\item H$_2$) There is a significant correlation between Team Maturity and Continuous Integration \& Testing (cit).

  $H_{2}:\rho_{tm,cit} \neq 0$ 
\item H$_3$) There is a significant correlation between Team Maturity and Stand-Up Meetings (sum).

$H_{3}:\rho_{tm,sum} \neq 0$ 
\item H$_4$) There is a significant correlation between Team Maturity and Customer Acceptance Tests (cat).

$H_{4}:\rho_{tm,cat} \neq 0$ 
\item H$_5$) There is a significant correlation between Team Maturity and Retrospectives (r).

 $H_{5}:\rho_{tm,r} \neq 0$ 
\item H$_6$) There is a significant correlation between Team Maturity and Iteration Planning (ip).

  $H_{6}:\rho_{tm,ip} \neq 0$ 
\item H$_7$) There is a significant correlation between Team Maturity and Customer Access (ca).

$H_{7}:\rho_{tm,ca} \neq 0$ 
\end{itemize}

Where: $\rho$ = Correlation Coefficient

\section{Method}\label{method}

\subsection{Subject selection and data collection}
This study was carried out at seven organizations (see Table~\ref{fig:companies2} for an overview). Three companies and one university from Brazil, and three companies from Sweden participated in the study. All the participating companies stated they had been using an agile approach in their software development for at least three years but with teams of different maturity levels in that process. The companies were found through our research network and the second author was teaching the course with the student sample. Since the student sample did not have lower maturity than practitioners on any measurements (see below), we opted to analyze all data in conjunction. The first author did all the data collection across all the participating organizations and did not know any of the contact persons before this research started. The data collection took place between February 2015 and June 2016.

The Brazilian sub-sample contains data from IT departments at one large on-line media and social networking enterprise with around 5,000 employees, one smaller software life-cycle consultancy company with around 35 employees, and one company that offers programming courses to individuals and companies with around 100 employees. We also collected data from the University of S\~ao Paulo including software engineering students enrolled in an agile software development project course where they developed software in small groups using an agile approach. We collected 132 responses from 22 groups in total. One response sheet lacked information of what team that person belonged to. The 22 groups had a total number of 142 group members that received the surveys via their managers but the replies were anonymous. The response rate was 93\% for the Brazilian sub-sample.

The Swedish sub-sample consisted of twelve groups across three large organizations: one being a multinational networking and telecommunications equipment and services company (around 115,000 employees), the second an aerospace and defense company (around 14,000 employees), and one automotive parts manufacturing company (around 160,000 employees). The twelve participating teams consisted of 68 agile team members and the survey was initially distributed to 77 people. The response rate for the Swedish sub-sample was therefore 88\%.

In total, the survey was distributed to 219 agile team members and 200 responded, hence an overall response rate of 91\%. This high response rate was due to the fact that the surveys were distributed in paper form during a special session on site at each company. A total of 34 agile teams participated in this study, and the team sizes ranged from 3 to 11 members for the participating teams.

Since we wanted to look at connections between averages on a team-level, we did not differentiate between different types of team or roles in the organizations. Nor did we look at where they were placed in the organization or the content of their software development tasks, or at differences between the two countries due to the small Swedish sub-sample. However, having data from two different continents strengthens the external validity.

\begin{table}
\footnotesize
\renewcommand{\arraystretch}{1}
\caption{Participant information.}
\label{fig:companies2}
\begin{tabular}{p{15mm}p{30mm}p{30mm}p{30mm}p{20mm}p{20mm}}

Organizations \bfseries & Business & Number of employees & Participating teams & Participating individuals. & Country\\
\hline
Company $1$ & On-line media  & 5,000 & 3&22&Brazil\\

Company $2$ & Software consultancy & 35 &8 &57&Brazil\\

Company $3$  & On-line training platform &100 & 3&11&Brazil\\

University students  & Agile course & 43 (enrolled students attending) &7& 43 &Brazil \\

Company $4$  & Telecommunications & 115,000 &4 & 21&Sweden\\

Company $5$ & Defense  & 14,000 & 4&24&Sweden\\

Company $6$  & Parts manufacturing  & 160,000 &5 & 22&Sweden\\
\hline
&   & Total  &34 & 200\\
\hline
\end{tabular}
\end{table}

\subsection{Operationalizations of Group Maturity and a subset of Agile Practices}
The Scale IV of the English Group Development Questionnaire \cite{wheelan} (or GDQ) was used to assess maturity of groups in this study. The 15-item Scale IV of the GDQ is rated on a Likert scale from 1 (Never true of this group) to 5 (always true of this group), and measures the work and productivity aspects of a work-group. Due to copyright reasons only three example items can be included here: 1) The group gets, gives, and uses feedback about its effectiveness and productivity. 2) The group acts on its decisions. 3) This group encourages high performance and quality work.

In order to get a full profile of the group development, the whole 60-item survey for the four different stages needs to be used. However, immature groups will get lower scores and mature groups will get higher scores on Scale IV. Also, previous group development studies have shown the mean of Scale IV to be significantly correlated to other external measurements of effectiveness (e.g.\ \cite{wheelan1998,wheelan1999,wheelan2005,wheelan20032,jacobsson2014links}). The full GDQ was also used in a sub-sample of the data and tested the hypothesis that Stage II (Counter-dependency and fight) would be negatively connected to the agile practices, which was confirmed \cite{gren2017links}.

To measure agile practices we used the survey developed by So and Scholl \cite{so}, which was also in English. As mentioned, it is the only survey we have found that is validated through a factor and reliability analysis ($N=227$). The measurement comprises items and factors for the seven agile practices \emph{iterative development} (7 items), \emph{continuous integration and testing} (9 items), \emph{stand-up meetings} (5 items), \emph{customer acceptance tests} (5 items), \emph{retrospectives} (6 items), \emph{iteration planning} (7 items), and \emph{customer access} (4 items). All items for each scale can be found in So and Scholl \cite{so}, and in their study, the scales had Cronbach's $\alpha$s ranging from .82 (good) to .95 (excellent). In our present study, the $\alpha$s were between .72 (acceptable) and .97 (excellent), and the lowest value was for the practice Stand-up meetings in both cases. The scale for assessing Cronbach's $\alpha$s is that less than 0.5 is unacceptable, between 0.5 and 0.6 is poor, between 0.6 and 0.7 is questionable, between 0.7 and 0.8 is acceptable, between 0.8 and 0.9 is good, and more than 0.9 is redundant and not preferred, i.e.\ too many similar questions \cite{streiner2003starting}. The seven scales are replied to on a Likert scale from 1 (never) to 7 (always). We have already presented two example items from each practice in Section~\ref{agileprac}.



\subsection{Data Analysis}
In order to analyze the connections between the agile practices and group maturity we ran separate and bivariate correlation analyses on all seven agile variables and GDQ IV, using the mean of all the sums for each scale. For each scale (set of items), like GDQ IV, Iterative Development, etc., instead of taking the mean value for all the items in the scale as a measurement for that individual's score, we sum up the item answers instead. The mean of the sums is then the all the individuals' sums in each team divided by the number of individuals. To clarify, for each team, we calculate $\frac{\sum_{j=1}^{m}(\sum_{i=1}^{n}item_{i})}{m}$ where $n$= the number of items in a scale (e.g.\ team maturity), and $m$ is the numbers of team members. The reason for using sums of item responses for each scale was due to the fact that all the previous studies on the GDQ uses the sums instead of only the means, which allows a comparison between studies. The Pearson Product-Moment correlation coefficient was used as a measurement of correlations between mean values of the 34 groups on all scales, except \emph{iteration planning} and \emph{customer access}, since the Shapiro-Wilk tests of normality in data were significant for these variables ($Test$ $Statistic = .850$, $p<.000$, and $Test$ $Statistic = .924$, $p=.021$). For these two variables we instead ran Spearman's $\rho$ correlation analyses. Since we did not know the direction of the correlations we opted to use two-tailed tests. To assess the size of the effects we compared the coefficients to the guidelines suggested by Cohen \cite{coheneffect}.

We also ran Mann-Whitney U tests to see if there were any differences between countries and between students and practitioners. The results from the Mann-Whitney U tests to see if there were any differences between countries and between students and practitioners, we found a significant difference between Sweden and Brazil on the \emph{customer access} practice ($W$ = $215$, $p=.003$). The Swedish teams had less customer access which was most likely due to the sizes of the participating Swedish companies. Regarding differences between students and practitioners we found that students had significantly higher on \emph{customer access} ($W$ = $26$, $p=.004$) and \emph{iteration planning} ($W$ = $29$, $p=.004$) than practitioners. There were no significant differences on any other agile practices nor in team maturity across countries or type of respondent (student or practitioner teams). The fact that students scored higher on two of the agile practices led us to assume that they were at least as mature in using them as the practitioners, hence we kept them in the analysis.

\section{Results}\label{sec:res}

\begin{table}
\renewcommand{\arraystretch}{1.5}
\caption{Descriptive statistics for the included variables.}
\label{fig:descr}
\centering
\begin{tabular}{lll}
\hline
\bfseries Measure &Mean of the teams' mean sums & Standard Deviation of the  teams' mean sums\\
\hline
1. GDQ Scale IV 					&   60.0			&   4.4 \\
2. Iterative Development 				&  37.4 	&  4.1    \\
3. Continuous Integration \& Testing 	  &  39.0  &   7.0\\
4. Stand-Up Meetings 				 & 26.7 & 3.5\\
5. Customer Acceptance Tests			&19.2 &     5.5   \\
6. Retrospectives 					&  29.6 	&    6.0  \\
7. Iteration Planning 				&   38.4		 &  5.1   \\
8. Customer Access 				&  19.7 &  5.3\\
\end{tabular}
\end{table}

The mean of all the mean sums and their standard deviations for the 34 groups on the used scales are shown in Table~\ref{fig:descr}. As can be seen in Table~\ref{fig:corrmatrixapa}, more mature teams make the intended use of the following practices ($p<.01$): (2) \emph{iterative development} (.647**) which is a strong correlation (or a large effect), and (6) \emph{retrospectives} (.490**), which is a medium correlation\slash effect but very close to a large effect of .5. Thus, we reject the null hypothesis in favor of $H_{1}$ and $H_{5}$ and fail to reject the other hypotheses. The size of the effects were compared to the effect size guidelines as suggested by Cohen \cite{coheneffect}. This means that a higher group maturity is connected to a better implementation of \emph{iterative development} and \emph{retrospectives}. It also means that if the group is less mature, i.e.\ obtained lower values on GDQ Scale IV, these two agile practices will probably not work as intended. However, since a bivariate correlation analysis was conducted, we do not know the direction of the effect, and these two agile practices might instead lead the group to mature.


\begin{table}
\renewcommand{\arraystretch}{1.5}
\caption{Pearson Correlations of Variables ($N=34$ teams).}
\label{fig:corrmatrixapa}
\centering
\begin{tabular}{lllllll}
\hline
\bfseries Measure & 1 & 2 & 3 &4 & 5 & 6  \\
\hline
1. GDQ Scale IV 					&  1 			&   &  &  &  &   \\
2. Iterative Development 				&   .647**		&   1 &  &   &  &    \\
3. Continuous Integration \& Testing 	&   .314 		&   .569** & 1  &  &    &   \\
4. Stand-Up Meetings 				& .155 		& .450** & .184    & 1    &  & \\
5. Customer Acceptance Tests			&   -.040 		&   .166 & .310    & .367*    &  1    &      \\
6. Retrospectives 					&   .490** 	&   .511** & .517**    &  .234     & -.316     & 1    \\
Notes: *$p<.05$, **$p<.01 $ 
\end{tabular}
\end{table}

The result of the two non-parametric correlation analyses are shown in Table~\ref{tb:cor:cha}. The results show that none of the variables were significantly correlated to GDQ Scale IV. Another interesting aspect of the correlations is that all we can confirm So's \cite{so2010making} result that \emph{iterative development} seems to be at the core of agile software development since \emph{continuous integration \& testing}, \emph{stand-up meetings}, and \emph{retrospective} were all highly correlated to this variable. 

\begin{table}
\renewcommand{\arraystretch}{1.5}
\caption{Spearman Correlations of Variables ($N=34$ teams).}
\label{tb:cor:cha}
\centering
\begin{tabular}{llll}
\hline
\bfseries Measure & 1 & 2 & 3   \\
\hline
1. GDQ Scale IV 					&  1 			&   &  \\
7. Iteration Planning 				&   .324		&   1 &     \\
8. Customer Access 	&   .012 		&   .263 & 1  \\
\end{tabular}
\end{table}

\section{Discussion}\label{disc}
The results of this study show that there are strong connections between a group's maturity, as represented by the \emph{GDQ Scale IV} measurement and two concrete agile practices. The significant correlations found between \emph{iterative development} and \emph{retrospectives}, and \emph{GDQ Scale IV} give us insight into what agile practices are connected to the group maturity level in order to function in the intended way. In addition, the agile practice of developing software iteratively is the core of agile software development, according to So \cite{so2010making}, which we have confirmed in our result. This also means that the behavior connected to the other practices \emph{continuous integration and testing}, \emph{stand-up meetings}, \emph{customer acceptance tests}, \emph{iteration planning}, and \emph{customer access} could work well even in the early phases of new agile teams, since we fail to reject these null hypotheses, but they are still linked to the iterative development of software. We need more studies on these variables to conclude on their seemingly quite complex inter-correlations and how they interplay causally. Even though we want to highlight the fact that failing to reject the null hypothesis does not lead to the conclusion that it is true. Regardless of this fact, the items measured in the practices \emph{iterative development} and \emph{retrospectives}, as suggested by So and Scholl \cite{so} are in connection to higher requirements regarding the group maturity as compared to the other practices measured, by asking about, for example, the degree of active discussion, areas of improvement, and evaluation of solutions instead of more technical or process-related practices. The variable with almost a significant effect was \emph{iteration planning}, which also includes items more related to how active group members are in the team. In addition to the \emph{retrospectives}, the \emph{stand-up meetings} are also a synchronization of team members and we would not have been surprised if this practice was also correlated to group maturity. However, the way this practice is measured in the survey by So and Scholl \cite{so} we see that the items regarding \emph{stand-up meetings} are shallower then for the \emph{retrospectives} for example with question on if the meeting was ``to the point'' etc.\ without details in the expected behavior. Maybe these items do not fully reflect the intended behavior in a stand-up meeting, but this remains to be explored. 

Since a bivariate correlation analysis was applied, we do not know the direction of the effect, i.e.\ which variables are the cause and effect, and therefore team agility might be dependent on group maturity or the agile practices might lead to the group maturing. In order to investigate the causality between these measurements, longitudinal studies are needed where the effect is preceded by the cause. The results must also be consistent between studies, and eventually a theory is needed that describes the connections between the constructs, which then can be tested empirically.

The results of this study therefore contribute to our understanding of why a transition to a more agile approach is more difficult for some teams at certain points in time. One could argue that most suggested processes of effective teamwork assume mature teams \cite{combs2006much}, but it is evident that there is more of a team focus in the agile approach compared to others \cite{melnik}, i.e.\ groups dynamics is more important to understand in agile team than traditional ones. In summary, this provides an answer to the question (the second research objective) of which (if any) of the selected agile practices are connected to team maturity in order to function as intended.

\subsection{Implications for Psychology}\label{sec:1.33}

This study links a more well-known concept within organizational psychology, namely group development, to new work processes that evolved in practice in a new type of industry, namely software development. Because the group development measurement was linked to two core practices in agile development, this means that a group of higher maturity can then be more responsive to change, which is not explicitly described in group development theory \cite{wheelandev}.

\subsection{Practical implications}\label{sec:1.3}
Our results imply that including aspects of group maturity could increase the understanding and predictability of agile implementations. We would like to highlight here that a low team maturity could as well be due to external factors in the company ecosystem outside of the team's control. This aspect is also highlighted below under Organizational Support. The result above entails that leaders and managers could be helped by preparing for the disparity of agile practices in use and accept this as a natural part of the group's development. Practitioners should be aware of that the group could make different use of the agile practices when less mature and only leverage these practices when the group has developed further, which is something we have already seen indications of in software engineering research \cite{lehtinen2017recurring}. Even if some practices will be used differently in the first compared to the latter stages of group development, this does not imply that these practices should not be implemented in the beginning. Rather, such practices are most likely needed for the group to develop since e.g.\ \emph{retrospectives} give the group a forum to reflect on group work, which is needed in all stages. However, the content will differ since the group has different issues to work on from a group developmental perspective. To clarify, the correlations shown in this study connects the two constructs of agility and team maturity, which implies that agile teams can use the group development model and its appended interventions in order to become more responsive to change in their software development. 

\paragraph{How to create an effective team}\label{sub:howto}
In order to give more concrete recommendations of what agile teams can do if wanting to improve their team maturity, we have summarized the twelve keys to productivity for groups as presented by Wheelan \cite{wheelan2012}. These points apply to teams in general, but based on the results of this current study, working on these aspects could increase the agility of teams (even if the causality aspect needs further investigation). Fore more details and references, see Wheelan \cite{wheelan2012}.

\begin{itemize}
\item Goals
\item Roles
\item Interdependence
\item Leadership
\item Communication and Feedback
\item Discussion, Decision-Making, and Planning
\item Implementation and Evaluation
\item Norms and Individual Differences
\item Structure
\item Cooperation and Conflict Management
\item The Shared Responsibility 
\item Organizational Support
\end{itemize}

These aspects are very similar to the theory of Process Consultation created by Schein \cite{schein1988} and backed up by an extensive study by Senior and Swailes \cite{senior2004} on what team building activities actually work. Working with any of these aspects as an external resource can be defined as an intervention by a consultant, and in the consultancy theory, these aspects have been known for decades \cite{lowman}. We will now go through them one by one and briefly explain what needs to be done.

\paragraph{Goals}
That members are clear about team goals is the single most important part of high performing teams. Teams often seem to know their groups but team members usually have different ideas about what accomplishment acquires, and even if they use the same words they can have a different meaning to different people. The only way to make sure members understand the goal is to discuss it thoroughly. Without being clear about group goals, there is no meaning trying to accomplish them. However, knowing the goal is not enough, in high performance teams, members actually agree with the team goals. That highlights the importance of goals being relevant and preferably important, reasonable, and attainable to benefit both the team and the organization \cite{wheelan2012}. 

\paragraph{Roles}
After the goal is defined the group can get organized and decide what needs to be done and who does what. The most important thing is that each member really knows what their role is, independently of if they volunteered for the role or not, i.e., both the expectations and the process need to be clear.This clarity of roles and process refer to clarity within the teams, and does not imply the same roles and processes across the organization. Secondly, the person taking on the role must have the ability and skills that is needed for the tasks. Thirdly, the whole group must agree on all the different roles since they are a part of a whole solution \cite{wheelan2012}. 

\paragraph{Interdependence}
In high performance teams, the tasks demand that members work together as a unit or in subgroups to reach the goal. This means that the group should not be too big \cite{wheelan2009}. The goal should also not be possible to reach without team work \cite{wheelan2012}.

\paragraph{Leadership}
In a successful team, the leader changes her leadership style according to emerging group needs. The leadership style also adapts to the group development stage. One important aspect of leadership is that focus needs to be drawn from the leader to the group. It is important to connect the aspect of culture as being the other side of the same coin as leadership, with the success of a group. All members of a group are an essential part of its success \cite{wheelan2012}.

\paragraph{Communication and Feedback}
One of the main characteristics of a high performance team is that it has an open communication structure, which allows all members to actively participate. This means that no matter which race, sex, ethnicity, profession, etc.\ a person has, their idea is still heard. This is easier said than done because of a cultural legacy in society which promotes and sustains inequality to keep the already existing power structure \citep{acker}. High performing teams also get constant feedback on their productivity and effectiveness both internally and from external resources, and use this feedback to make improvements to the group work. That is, it is not enough getting the feedback, but utilizing it to make changes is essential. Constructive feedback is also a key aspect, since it promotes improvement and individual development. This type of feedback needs to be strictly connected to making the group even more effective. If the feedback is unrelated to improving group work, it should not be given \cite{wheelan2012}. 

\paragraph{Discussion, Decision-Making, and Planning}
The best teams spend time planning how they will solve problems and make decisions. This means that they know how decisions are supposed to be made, before they make them. It does not matter if the decision-making strategy is consensus or majority, etc., it is only important that the rules of engagement are defined beforehand. The strategy itself also needs to be generally accepted and effective. Another aspect is that these teams spend time defining and discussing the problems they must solve. These teams actually spend more time preparing for future work compared to other teams. This takes more time in the beginning, but this time is easily made up for when tiresome strategic disagreement can be avoided every time a decision needs to me made \cite{wheelan2012}.

\paragraph{Implementation and Evaluation}
To get a well-functioning team, you need to implement the solutions and decisions made. This means that it is necessary to follow up on decisions and evaluate their success. Decisions can be unfavorable, but in high performance teams they are altered when this information is available to the team \cite{wheelan2012}. 

\paragraph{Norms and Individual Differences}
Group norms are utterly important for effectiveness. This result has also been replicated in software engineering research \cite{teh}. High performance teams have norms that encourage high performance, quality, and success. The aspect of what is expected of the team is key. If the group is not expected to perform, the probability of success is lower. Successful teams also accept differences in people as long as their behavior helps task accomplishment \cite{roberge}.

\paragraph{Structure}
The structure of the team is important for three reasons. The team members must all contribute, and therefore, a successful team only consists of the smallest number of members necessary to reach the group goal. The group must also allow subgroups to form to work on smaller chores. These subgroups are not seen as a threat to the group, but as necessary and valued for their contribution to the team. One aspect often not mentioned are the resources available to the team. A high performing team is organized but also given enough time together in order to develop and maintain this high performance \cite{wheelan2012}. 

\paragraph{Cooperation and Conflict Management}
A high performing team consists of cooperative members and has great cohesion. However, groups can cooperate in order to not work. High performing teams have conflict frequently, but these periods are short and task-oriented. The key is that these teams have well-working conflict management strategies \cite{wheelan2012}. How to manage conflict effectively has been known for quite some time \cite{thomas2006}.

\paragraph{The Shared Responsibility}\label{sub:shared}
A leader (informal or formal) can affect a group's development vastly, but is not solely responsible for the success of the group. That is why it is so important to recruit people with professional goals that are aligned with the organizational goals. The individual also needs room for self-fulfillment within the scope of the employment \cite{maslow}. Every team member is responsible and can make a difference in how well-functioning a group becomes by doing what they can to help the group develop \cite{wheelan2012}. 

\paragraph{Organizational Support}\label{sub:organizationalsupport}
In order to get high performing teams in an organization, the organization itself needs to give the teams the right support. Groups have difficulty in performing well if not placed in a favorable organizational culture. The organization needs to clarify the mission, support innovation, expect success, value superior quality and service, pay attention to detail, value team recommendations, set clear expectations for group output, quality, timing, and pacing, and reward teamwork rather than individual performance \cite{wheelan2012}. 

It is also necessary to give the groups what they need to in order for them to do their best. Some of these aspects have been mentioned before and have to do with goals and tasks, and continuous learning. It is also important that the groups get the human resources required to deliver, as well as technical. Groups should have a defined work area \citep{wheelan2012}. 

As mentioned before, the effectiveness of a team is dependent on all the group members' joint effort. Members need to be recruited to the project based on their skills needed for group success. It is also important to train members in group participation. However, all training needs to be based on research evidence and proven to work. There are plenty of companies offering group development that has no proven effect \cite{klein}. According to Sundstrom et al.\ \cite{sundstrom1990} and Guzzo et al.\ \cite{guzzo1985}, the most effective interventions are goal setting and feedback that includes attention to group development issues adapted to the current group stage they are in.

Another interesting aspect of organizational support is that it should not be too extensive. If a group has too many helpers, the group members might not learn to be effective and productive on their own. Groups need feedback so that they can help themselves. The groups need enough autonomy to get the work done, but at the same time stay connected to the organization, and regularly have reviews on the organizational support \cite{wheelan2012}.

In summary, the above recommendations from group psychology and, specifically, group development theory and practice are not dissimilar to the descriptions of agile teams. This is no surprise, but to work on team maturity to get more agile teams is not always well known to agile practitioners.

Another direct consequence of focusing on team maturity when building agile teams, is to not see software developers as technical resources only that can be moved between different teams without any negative psychological effects, especially since stable teams over time are more likely to have matured \cite{wheelan2003}. Instead, it is important to take the group development into account when building and maintaining software development teams.

\section{Validity Threats}\label{vts}
In this study, we used the most validated agility measurement survey we could find, but of course, the operationalization of agility as a construct is a great threat to construct validity of the measurement used. We believe the measurement captures the important aspects of what is meant by agile behavior, but we cannot be entirely sure of this without further studies on the content validity. Earlier research has also pointed out the difficulty of measuring agility (e.g.\ Gren et al.\ \cite{grenjss}). Another threat is that we assumed that the participating organizations were actually using agile practices correctly, which might not have been the case. We did not observe the teams using the practices and only confirmed their use through our company contacts. However, the surveys used to measure the agile practices do not make sense without some introduction of the practices to the terms, and we also tried to capture the degree of their implementation, which makes the output relevant for a correlation analysis.  

Another threat is that we only looked at factors internal to the teams. Team maturity is, as we have seen, also dependent on external factors to the teams themselves that then are reflected in less maturity. We only know how these agile practices are perceived in connection to a self-assessment of the team members' own teams, hence we do not know what external factors that could affect both factors.

We did not look closely at the boundaries of the teams or the different roles often included in software development teams (like testers, developers, architects, leaders, etc.). It would be interesting to look at role composition in the software development context in connection to group development. Another threat to the construct validity is the fact that all questionnaires were filled out in English and not in the native languages of the participants, i.e.\ Portuguese or Swedish. This could mean that the participants consistently misinterpreted some words or expressions. However, since we collected data from two different countries with two very different native languages that potential systematic error is decreased. 

We would like to state, again, that broad generalizations of our result should be done with care, due to our smaller sample size. In addition, we employed a cross-sectional research design, which gives us difficulty in drawing any conclusions on causal relationships between our two measurements.

\section{Conclusion and Future Work}\label{conc}

In conclusion, the purpose of this study was to investigate the relationship between group maturity and agile software development practices. The result shows that two of the core agile practices were correlated to group development maturity, i.e.\ higher levels of group maturity were connected to better use of the agile practices \emph{iterative development} and \emph{retrospectives}. Despite our small sample size of 34 groups, we believe our study shows an indication of how some agile practices are connected to group maturity, and through these results, contributes to the emerging new research field of connecting social psychology to software engineering.

For future work, we particularly suggest more studies on possible mediators and moderator in the connection between agility and team maturity. There seem to be many confounding factors that need carefully planned survey studies or experiments. We also suggest the use of qualitative data since such data are very rich and can unveil both causal relationships and more confounding factors. 

In this study, we found that students obtained higher scores on the practices \emph{customer access} and \emph{iteration planning}. The first one could be due to more of the participating companies being larger, but an investigation into why students scored higher on these practices would be interesting to conduct.

Future research should also include larger sample size in order to draw conclusions for the general population of agile software development teams, and investigate these concepts over time in longitudinal studies, as mentioned above. For example, could we increase agility by helping the group in its group dynamic development? Or, does an agile software development process lead to the group maturing? The previously mentioned confounding factors could also be on an organizational level, which also need to be a part of the data in order to fully understand the dynamics of agile teams and what it is affected by.

\section*{Acknowledgements}
We would like to thank all the participants and their companies who helped making the data collection possible for this research. We would also like to thank Christoph Stettina for interesting discussions during the execution and analysis of this study.
\bibliographystyle{model5-names}
\bibliography{references}
\newpage

\end{document}